\documentclass[conference]{IEEEtran}
\renewcommand{\baselinestretch}{1}
\renewcommand{\baselinestretch}{0.98}
\usepackage{cite}
\usepackage{amsmath,amssymb,amsfonts}
\usepackage{graphicx}
\usepackage{textcomp}
\usepackage{xcolor}
\usepackage{url}
\usepackage[nobiblatex]{xurl}
\usepackage{multirow}
\usepackage{longtable}
\usepackage{tabularx} 
\usepackage{booktabs}
\usepackage{array}
\usepackage{placeins}
\usepackage{enumitem}
\usepackage{algorithm}
\usepackage{algpseudocode}

\renewcommand{\baselinestretch}{0.98}

\def\BibTeX{{\rm B\kern-.05em{\sc i\kern-.025em b}\kern-.08em
    T\kern-.1667em\lower.7ex\hbox{E}\kern-.125emX}}
\begin{document}

\title{AI-Enabled Data-driven Intelligence for Spectrum Demand Estimation\\}

\author{\IEEEauthorblockN{Colin Brown\IEEEauthorrefmark{1}\IEEEauthorrefmark{2}, Mohamad Alkadamani\IEEEauthorrefmark{2}\IEEEauthorrefmark{3}, and Halim Yanikomeroglu\IEEEauthorrefmark{3}}
\IEEEauthorblockA{\IEEEauthorrefmark{2}Communications Research Centre, Ottawa, Ontario, Canada}
\IEEEauthorblockA{\IEEEauthorrefmark{3}Carleton University, Ottawa, Ontario, Canada}
\IEEEauthorblockA{\IEEEauthorrefmark{1}Corresponding Email: colin.brown@ised-isde.gc.ca}
}

\maketitle

\begin{abstract}
Accurately forecasting spectrum demand is a key component for efficient spectrum resource allocation and management. With the rapid growth in demand for wireless services, mobile network operators and regulators face increasing challenges in ensuring adequate spectrum availability. This paper presents a data-driven approach leveraging artificial intelligence (AI) and machine learning (ML) to estimate and manage spectrum demand. The approach uses multiple proxies of spectrum demand, drawing from site license data and derived from crowdsourced data. These proxies are validated against real-world mobile network traffic data to ensure reliability, achieving an R$^2$ value of 0.89 for an enhanced proxy. The proposed ML models are tested and validated across five major Canadian cities, demonstrating their generalizability and robustness. These contributions assist spectrum regulators in dynamic spectrum planning, enabling better resource allocation and policy adjustments to meet future network demands.
\end{abstract}



\section{Introduction}\label{sec:introduction}

Network traffic is expected to grow exponentially due to advancements in smart devices, the proliferation of the internet of things (IoT) and increased usage of cloud computing. This surge is particularly notable in wireless networks, driven by the advent of 6G standards, new spectrum users, and the growing demand for wireless access. This growth places increasing pressure on mobile network operators (MNOs) and regulators to manage spectrum resources efficiently and ensure a sufficient supply to support future spectrum demands.

Spectrum demand within a wireless network is typically monitored by the respective mobile operators, making it difficult for spectrum regulators to  directly observe and thus make sufficient plans to regulate the spectrum. However, understanding and modeling current spectrum demand, or proxies thereof, is highly advantageous for regulators and other spectrum stakeholders. These capabilities aid in planning the release of new spectrum, identifying geographical areas with an under- and over-supply of spectrum, and adapting policies and licensing schemes to alleviate future demand pressures. 

Data-driven approaches enhanced by artificial intelligence (AI) technologies are widely recognized for their potential in providing intelligence for spectrum management and facilitating spectrum demand estimates~\cite{Saruthirathanaworakun2024,FCC_data_driven}. However, estimating spectrum demand presents several challenges, including selecting and validating appropriate proxy data, which are reliable substitutes for actual traffic demand data when direct measurements are unavailable, that accurately represent spectrum requirements. Additionally, obtaining geospatial data to understand the variability of spectrum demand across different geographical areas is necessary. Finally, ensuring that AI models can predict spectrum demand and be generalized across various scenarios is crucial.

\begin{figure*}
\centering
\includegraphics[width=0.9\linewidth]{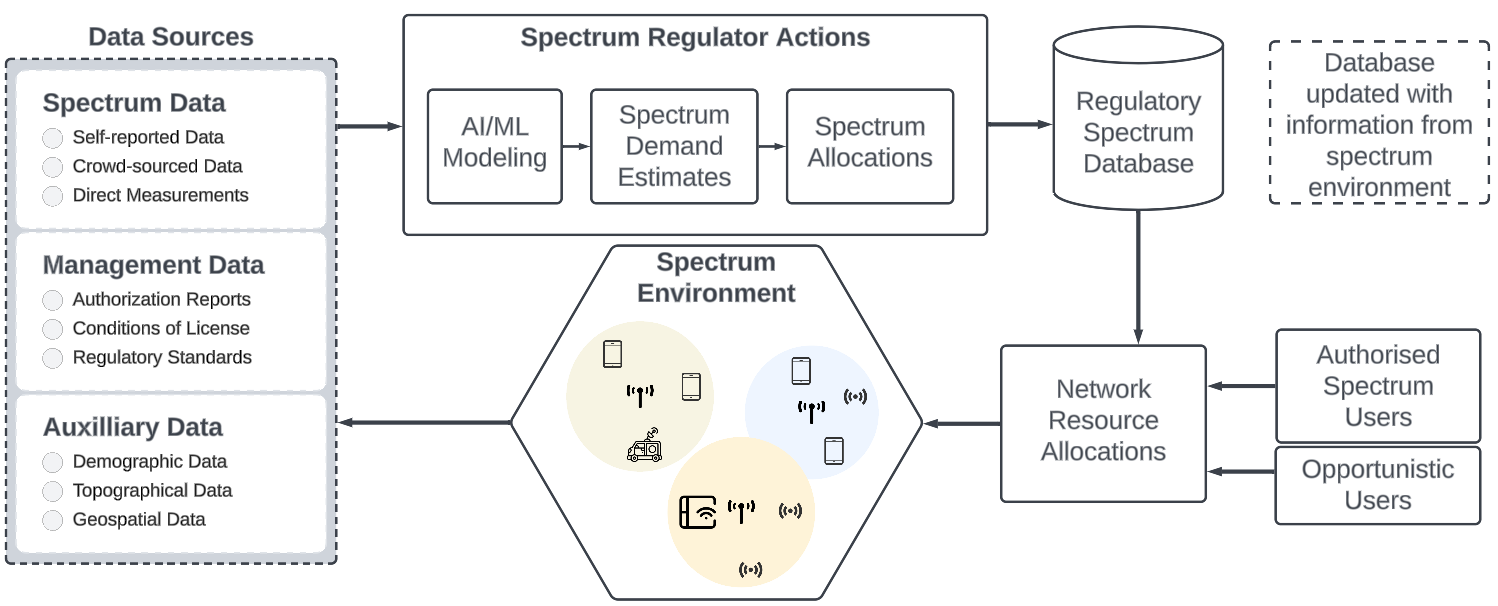}
\caption{Integration of AI-enabled spectrum demand modeling within the broader spectrum planning and resource allocation framework.}
\label{fig:spectrum_regulator_actions}
\end{figure*}

Figure~\ref{fig:spectrum_regulator_actions} illustrates the integration of spectrum demand estimation into the regulatory framework of spectrum planning and allocation. On the left side, various datasets are shown, including Spectrum Data (self-reported, crowd-sourced, and direct measurements), Management Data (authorization reports, conditions of licenses, and regulatory standards), and Auxiliary Data (demographic, topographic, and geospatial data). These datasets are processed using AI-enabled modeling to generate Spectrum Demand Estimates, which inform Spectrum Allocations by regulatory bodies.

These allocations are subsequently recorded in a Regulatory Spectrum Database, ensuring decisions are data-driven. The Network Resource Allocation manages how the resources are allocated to Authorized Spectrum Users and Opportunistic Users prior to deployment in the Spectrum Environment. The feedback loop consists of data collected from the Spectrum Environment which can be used to evolve the information contained within the data sources and allows the spectrum regulatory actions to adapt to the network conditions and patterns of demand.

In this context, a data-driven approach using AI technologies is proposed for generating intelligence on spatial spectrum demand, with a particular focus on validating and enhancing various demand proxies and developing machine learning (ML) models that are widely applicable.

Previous work in data-driven spectrum demand modeling has provided a relatively good starting point for further advancements. For example, in~\cite{ParekhPimrc2023}, the authors explored a diverse set of geospatial data inputs alongside a proxy for spectrum demand based on the number of mobile towers deployed by operators. This study identified key factors impacting spectrum demand, offering valuable insights despite the limited number of test scenarios and the absence of validation for the proxy.  Building on this, the work in \cite{ParekhFNWF2023} improved the spectrum demand proxy with a focus on model interpretability, although ML model generalizability was not specifically addressed. In \cite{Alkadamani2024}, a demand proxy derived from spectrum license data was examined and validated with mobile network operator data in two urban areas in Canada. While effective, this approach could benefit from exploring multiple proxies and enhancing generalizability across different regions.

Given the prior work in this domain, the contributions of this paper are threefold. First, a methodology to estimate spectrum demand using a new proxy from crowdsourced data is presented. The results highlighting a strong spatial relationship with real-world download traffic data from an MNO. Second, an enhanced spectrum demand proxy is developed through the combination of spectrum site license and crowdsourced data. Lastly, the broad applicability of the resultant ML model is presented via model training and validation over five urban scenarios.

The remainder of the paper is structured as follows: Section~\ref{sec:proposed_methodology} provides an outline of the proposed methodology for spectrum demand estimation. Section~\ref{sec:experimental_setup} then provides more details on the validation and modeling approaches. Results and discussion are presented in Section \ref{sec:results_discussion}. Lastly, Section~\ref{sec:conclusion} concludes the paper with key findings.

\begin{figure}
\includegraphics[width=0.92\linewidth]{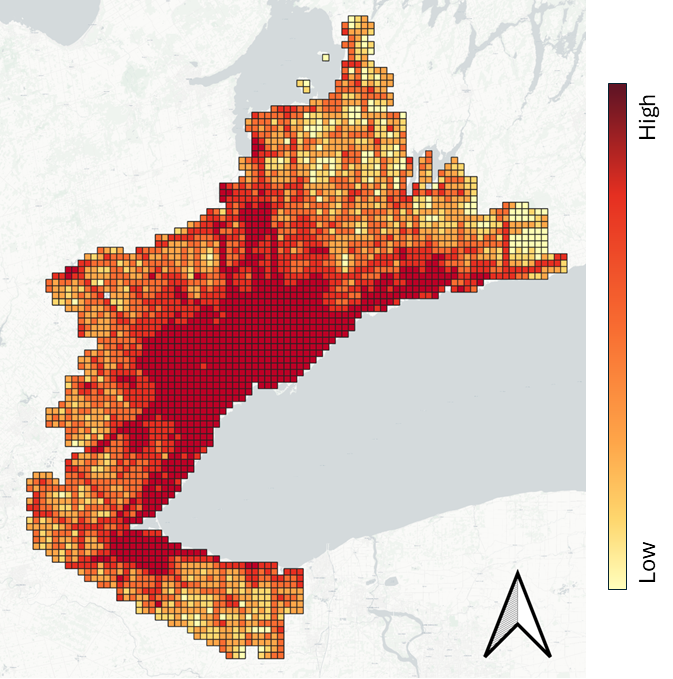}
\caption{Counts for the number of active user for a single MNO in Toronto taken from crowdsourced data and aggregated at grid tile level.}
\label{fig:heatmap_num_active_usr}
\end{figure}
\section{Proposed Methodology}
\label{sec:proposed_methodology}

Given the concentration of demand pressures in urban areas, this study emphasizes five major Canadian cities: Montreal, Ottawa, Toronto, Calgary, and Vancouver. Each city is segmented into a uniform grid system, with each grid tile measuring approximately $1.5\,\text{km} \times 1.5\,\text{km}$. This spatial grid framework facilitates the consistent spatial aggregation and examination of diverse input features and proxies.

An example of this is shown in Fig.~\ref{fig:heatmap_num_active_usr}, where a heatmap illustrates the count of active users, in Toronto, derived from crowdsourced data and aggregated at the grid level. In this example, geographical areas with relatively high counts of active users (darker shading) correspond to dense urban centers, while lighter areas indicate fewer counts in more suburban and rural regions. Additionally, infrastructure features such as the road networks are clearly observable, reflecting user mobility and commuting patterns in the area.

In the following section a methodology for estimating spectrum demand is presented.  The methodology consists of three main steps: constructing and validating proxies, integrating spatially processed features, and applying ML models while ensuring consistency in data splitting and evaluation.

\subsection{Proxy Development and Validation}
Three proxies are used to estimate spectrum demand, derived from two distinct data sources and spatially aggregated to the grid tiles for consistent analysis.

\begin{itemize}
    \item \textbf{Deployed Bandwidth Proxy (\(P_{\text{BW}}\)}): This proxy represents the total spectrum bandwidth deployed by MNOs across different regions, providing an estimate of available network capacity. It is extracted from self-reported site license data~\cite{ised_site}. Deployed bandwidth per grid tile is calculated by estimating each base station's coverage area using transmit parameters. Overlapping coverage areas are aggregated, and the deployed bandwidth is then mapped to each grid tile as follows:
    
    \begin{equation}\label{eq:proxy_bw}
    P_{\text{BW}, j} = \sum_{k \in \mathcal{C}_j} \text{BW}_k \cdot w_k,
    \end{equation}
    where \(\mathcal{C}_j\) represents the set of base stations covering grid tile \(j\), \(\text{BW}_k\) is the deployed bandwidth of base station \(k\), and \(w_k\) is a weighting factor accounting for the overlap fraction of the base station coverage area in grid tile \(j\).
    \item \textbf{Active Users Proxy (\(P_{\text{Users}}\)}): This proxy represents the density of active mobile users within each grid tile, reflecting real-world user activity patterns. It is generated using crowdsourced data, where various key performance indicators (KPIs), including user activity, are recorded daily. The number of unique users in each grid tile is calculated per day, and the total count is averaged over a specified time interval period:  
    \begin{equation}\label{eq:proxy_user}
    P_{\text{Users}, j} = \frac{1}{T} \sum_{t=1}^T U_{j,t},
    \end{equation}  
    where \(U_{j,t}\) represents the number of unique users in tile \(j\) on day \(t\), and \(T\) is the total number of days in the averaging period.

    \item \textbf{Combined Proxy (\(P_{\text{Combined}}\))}: To mitigate the limitations associated with individual proxies, a combined proxy is proposed, leveraging data from both self-reported deployed bandwidth and crowdsourced active user measurements. The rationale for combining these two data sources stems from their inherent strengths and weaknesses. \(P_{\text{BW}}\), while informative about infrastructure capacity, is self-reported by MNOs and may be subject to reporting errors, biases, or lags in reflecting actual spectrum demand. Conversely, \(P_{\text{Users}}\) offers insights into user distribution but may introduce biases in sparsely populated areas due to varying application adoption levels. To achieve a more balanced representation of spectrum demand, the proposed combined proxy integrates both metrics by weighting them based on their correlation with real-world traffic data:
    \begin{equation}
        P_{\text{Combined}, j} = \alpha_{\text{BW}} \cdot P_{\text{BW}, j} + \alpha_{\text{Users}} \cdot P_{\text{Users}, j},
    \end{equation}
    where \(\alpha_{\text{BW}} + \alpha_{\text{Users}} = 1\). The weights \(\alpha_{\text{BW}}\) and \(\alpha_{\text{Users}}\) are determined empirically to optimize alignment with traffic demand, ensuring that both infrastructure and user activity contribute meaningfully to the final proxy. 
    \end{itemize}

The proxies are evaluated and compared at two levels: (i) their ability to represent real-world traffic demand based on data from an MNO, and (ii) their performance in ML models to predict spatial spectrum demand.

\subsection{Feature Integration}

A diverse set of spatial features is aggregated into grid tiles to model the proxies accurately. For each grid tile \( j \), the feature vector is defined as \( \mathbf{X}_j = \{ x_1, x_2, \dots, x_k \} \), where \( x_i \) represents the \( i \)-th feature. These features, which serve as inputs to the model, include:

\begin{itemize}
\item \textbf{Demographic Data:} Population density, age distribution, and household composition. These demographic features are used to understand population dynamics and social structures within each grid tile, informing proxies related to human activity and societal trends.

\item \textbf{Economic Data:} Business counts and income levels. Economic features provide insights into the financial well-being and economic activity of each grid tile, which are critical for modeling relevant proxies.

\item \textbf{Physical Data:} Building coverage, road density, and infrastructure characteristics. These physical features describe the built environment and are essential for proxies related to urban planning, land use, and accessibility.

\item \textbf{Activity-Based Data:} Commuting patterns and traffic dynamics. These activity-based features capture the movement of people and vehicles, helping to model transportation demands and mobility-related proxies.
\end{itemize}




\subsection{Predictive Modeling Framework}

To assess the effectiveness of the developed proxies, a ML model is trained and tested using both individual proxies and the combined proxy. This approach enables a comparative analysis to determine which proxy best represents spectrum demand across different urban contexts.

A key aspect of this study is the evaluation of models across five cities. This ensures that the findings are not limited to a single geographic region but instead generalize across different urban environments with varying characteristics.


The models are evaluated using standard regression metrics, including the Coefficient of Determination (\(R^2\)), Mean Absolute Error (MAE), and Root Mean Squared Error (RMSE). These metrics collectively assess predictive accuracy and ensure that the model effectively captures variations in spectrum demand across different urban and suburban landscapes.



\section{Experimental Setup}
\label{sec:experimental_setup}

This section presents the experimental setup for both proxy validation and ML modeling. The validation analysis examines how well the developed proxies align with real-world mobile traffic data, while the ML modeling framework evaluates predictive performance across the five Canadian cities.

\subsection{Proxy Validation Setup}
The validation of proxies was conducted using real-world traffic data from the National Capital Region (NCR), Ottawa, collected from a 4G LTE network operated by a major MNO. The dataset captures network activity throughout 2018, including mobile traffic metrics such as download and upload throughput. The validation aimed to quantify how effectively the proxies represent actual spectrum demand.

\subsubsection{Traffic Data}
The traffic dataset contains hourly measurements of download and upload throughput (\(T_{j,t}\)), representing the total volume of transmitted data per hour. The dataset consists of 2,712 individual cell sites, processed as follows:
\begin{itemize}
    \item The coverage area of each cell was estimated using e-Hata propagation model~\cite{Okumura1968}, considering transmit power, frequency, antenna height, sector orientation (azimuth angle and beamwidth), and typical LTE receiver sensitivity.
    \item The computed coverage was mapped to grid tiles, with the fraction overlapping tile \(j\) denoted as \(C_{j,k}\). This fraction was used to proportionally allocate the cell's throughput:
    \begin{equation}
    C_{j,k} = f_{\text{e-Hata}}(P_{\text{TX},k}, f_k, h_{\text{ant},k}, \theta_k, \phi_k, S_{\text{RX}}),
    \end{equation}
    where \( P_{\text{TX},k} \) is transmit power, \( f_k \) is frequency, \( h_{\text{ant},k} \) is antenna height, \( \theta_k \) is azimuth angle, \( \phi_k \) is beamwidth, and \( S_{\text{RX}} \) is LTE receiver sensitivity.
    \item The total throughput for each tile \(j\) was computed by aggregating the contributions from all overlapping cells, weighted by their respective coverage fractions:
    \begin{equation}
    T_j = \sum_{k \in \mathcal{C}_j} T_k \cdot C_{j,k},
    \end{equation}
    where \( C_{j,k} \) is the coverage fraction of cell \( k \) within grid tile \( j \), \( T_k \) is the total throughput recorded for cell \( k \), and \( \mathcal{C}_j \) represents the set of cells covering grid tile \( j \).
\end{itemize}

\subsubsection{Proxy Computation}
The proxies were computed using the same methodology described in Section~\ref{sec:proposed_methodology}.

\paragraph{\(P_{\text{BW}}\)}
This proxy was derived from publicly available site license data for the same MNO operating in the study region during 2018. As sector-specific details were unavailable, each base station was assumed to have omnidirectional coverage per band. Coverage areas were estimated using the e-Hata model, and deployed bandwidth was mapped accordingly. The aggregated deployed bandwidth per grid tile was computed as per Eq.~\eqref{eq:proxy_bw}.

\paragraph{\(P_{\text{Users}}\)}
This proxy was generated using commercially available crowdsourced data collected in 2018. The dataset was obtained via SDKs embedded in mobile applications, capturing KPIs such as network connectivity, signal quality, and location. The number of unique users per day was computed for each grid tile as in Eq.~\eqref{eq:proxy_user}. 

\paragraph{\(P_{\text{Combined}}\)}
This proxy was derived by combining the two individual proxies (\(P_{\text{BW}}\) and \(P_{\text{Users}}\)) using empirically determined weights based on their correlation with the traffic data. Specifically, the weighting coefficients were set to \(\alpha_{\text{BW}} = 0.65\) and \(\alpha_{\text{Users}} = 0.35\) throughout.

To assess how well the proxies align with the real-world network activity, correlation analysis was performed using Ordinary Least Squares (OLS) regression. The primary metric used was the coefficient of determination (\(R^2\)).

To further assess the statistical significance of the regression models, the \(F\)-statistic was examined, which measures the explanatory power of the proxy relative to the total variance in network traffic data. A higher \(F\)-statistic indicates a stronger relationship between the proxy and actual network activity. Additionally, p-values were analyzed to determine statistical significance. A lower p-value (\(<0.05\)) suggests strong evidence that the proxy's correlation with real-world traffic data is not due to random variation.

\subsection{ML Modeling Setup}
Following the validation analysis, ML models were trained to predict spectrum demand using the three proxies as target variables. The goal was to evaluate the effectiveness of different proxies in a predictive modeling framework while ensuring generalizability across multiple urban regions.

\subsubsection{Data Preparation}
The study area was divided into grid tiles (\(1.5 \times 1.5\) km) covering urban cores and surrounding regions across the five Canadian cities analyzed in this study.

For each tile, the proxies were computed following the same methodology used in the validation analysis but extended across all cities. To ensure temporal consistency, all datasets were aligned with the year 2021, corresponding to the most recent Canadian census. This alignment was necessary as many statistical datasets, including demographic and economic indicators, are updated on a decadal basis, ensuring that all features and proxies correspond to the same timeframe.

\subsubsection{Spatial Processing}
To mitigate spatial autocorrelation and ensure a balanced distribution of the target variable, k-means clustering was used to divide each city into five distinct clusters, which were then used as folds for cross-validation. The clustering algorithm considered both geographic coordinates and the distribution of the target variable. This approach is particularly important in geospatial data, as neighboring tiles often exhibit similar characteristics. Without proper clustering, adjacent tiles could be assigned to both training and testing sets, leading to data leakage and overestimated model performance. In addition to clustering, another step was applied to further mitigate spatial autocorrelation by incorporating spatial lag features. This technique captures the influence of neighboring tiles by integrating their feature values, helping to account for spatial dependencies that may not be fully represented within an individual tile. The spatial lag feature is computed as $\text{Lag}(x_i)_j = \sum_{k \in \mathcal{N}(j)} w_{j,k} \cdot x_i,$ 
where \(\mathcal{N}(j)\) represents the neighboring grid tiles for tile \(j\), and \(w_{j,k}\) is a weight quantifying the spatial influence of tile \(k\) on tile \(j\).

\subsubsection{ML Framework}
The ML framework employed two models for predicting spectrum demand: a baseline linear regression model and an XGBoost (Extreme Gradient Boosting) model. The baseline model serves as a benchmark, relying on a single key predictor under the assumption of a linear relationship. In contrast, XGBoost is a more advanced tree-based model selected for its ability to capture non-linear relationships, hierarchical feature interactions, and its robustness to missing data. XGBoost optimizes the following regularized objective function:
\begin{equation}
L(\theta) = \sum_{j=1}^{N} l(y_j, \hat{y}_j) + \sum_{k=1}^{K} \Omega(f_k),
\end{equation}
where \( l(y_j, \hat{y}_j) \) represents the loss function (e.g., mean squared error), and \( \Omega(f_k) \) is a regularization term to control model complexity and prevent overfitting.

To evaluate the models' performance, the following metrics were used: \(R^2\), which measures the proportion of variance in the dependent variable that is predictable from the independent variable(s); Normalized RMSE (Root Mean Squared Error), computed as the square root of the average of the squared differences between predicted (\( \hat{P}_j \)) and actual (\( P_j \)) values, divided by the range of actual values (\( P_{\max} - P_{\min} \)), ensuring fair comparison across proxies with different scales; and Normalized MAE (Mean Absolute Error), which evaluates the average prediction errors while being less sensitive to large deviations, computed as the average of the absolute differences between predicted (\( \hat{P}_j \)) and actual (\( P_j \)) values, also divided by the range of actual values (\( P_{\max} - P_{\min} \)).
This structured approach validates proxies against real-world traffic data and integrates them into a predictive model, ensuring applicability across diverse urban environments.


\section{Results and Discussion}\label{sec:results_discussion}
\subsection{Proxy Validation Results}

\begin{figure}[ht] 
\centering 
\includegraphics[width=0.9\linewidth]{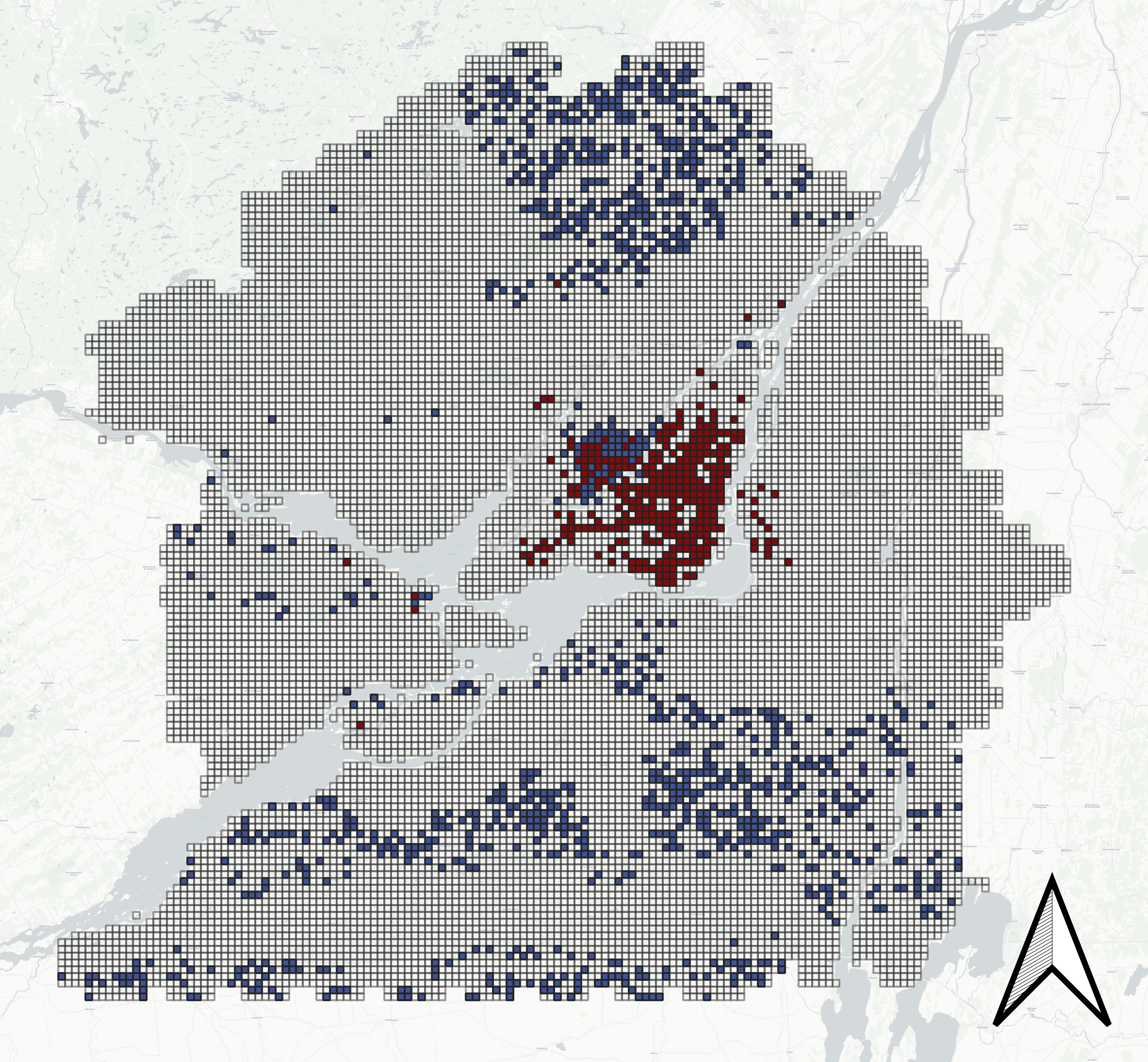} 
\caption{Geospatial distribution of proxy discrepancies in Montreal.} 
\label{fig:proxy_discrepancies} 
\end{figure}

A two-step analysis is conducted to assess the effectiveness of the proxies: first, examining their spatial behavior across urban and peripheral regions, and second, performing an OLS regression using real-world mobile traffic data from Ottawa.
The spatial analysis examines the rate of change for each proxy across different regions, comparing their relative variations. The results reveal that active user counts exhibit a steeper rate of change toward urban cores relative to deployed bandwidth, whereas deployed bandwidth increases more sharply in peripheral areas relative to active user counts. Figure~\ref{fig:proxy_discrepancies} illustrates this behavior in Montreal, where red areas indicate relatively higher user activity in downtown cores, while blue areas highlight greater bandwidth deployment in peripheral regions. This pattern does not imply lower bandwidth availability in red areas compared to blue areas but rather reflects differences in how each proxy scales spatially.
Several factors may contribute to this discrepancy. The deployed bandwidth may not always align with the actual demand, as infrastructure deployment often follows strategic planning rather than real-time user needs. As a result, network capacity in some areas may exceed immediate usage, while in others, demand may surpass available resources. Additionally, lower population density in peripheral regions may lead to fewer SDK-based measurements, potentially underrepresenting demand.

These spatial patterns were observed consistently across all five cities, confirming that the proxies exhibit systematic differences rather than random behavior.

To quantify these differences, OLS regression was performed using the mobile traffic data from Ottawa. The results, as shown in Table~\ref{tab:validation_metrics}, show that the Deployed Bandwidth Proxy achieves an \( R^2 \) value of 0.72, indicating a strong correlation with traffic data but with potential overestimations in underutilized infrastructure zones. The Active Users Proxy, with an \( R^2 \) of 0.64, captures end-user activity but may underrepresent demand in areas with sparse measurements. The Combined Proxy, which integrates both proxies, achieves the highest \( R^2 \) value of 0.85, providing a more comprehensive representation of spectrum demand.

Beyond \( R^2 \), the \( F \)-statistic values further reinforce the statistical significance of the proxies, with the Combined Proxy yielding the highest \( F \)-statistic of \( 1.33 \times 10^4 \), indicating strong explanatory power. Additionally, the p-values confirm statistical significance across all proxies. 

These results show that integrating both proxies mitigates individual biases, offering a balanced and generalizable view of spectrum demand. By combining infrastructure deployment and user activity patterns, the Combined Proxy better reflects actual network traffic
\begin{table}[ht]
\centering
\caption{Validation Metrics for Proxy Analysis}
\label{tab:validation_metrics}
\begin{tabular}{lccc}
\toprule
\textbf{Proxy}           & \textbf{\(R^2\)} & \textbf{\(F\)-Statistic} & \textbf{p-Value} \\
\midrule
Deployed Bandwidth (\(P_{\text{BW}}\)) & 0.72 & 4477 & \(< 0.001\) \\
Active Users (\(P_{\text{Users}}\))    & 0.64 & 3998 & \(< 0.001\) \\
Combined (\(P_{\text{Combined}}\))    & 0.85 & \(1.33 \times 10^4\) & \(< 0.001\) \\
\bottomrule
\end{tabular}
\end{table}

\subsection{Prediction Results}

To assess the effectiveness of different spectrum demand proxies in ML-based modeling, XGBoost regression models were trained using three proxy datasets alongside a baseline model. The baseline model relied solely on the strongest single predictor, daytime population—the distribution of population density during working hours—as a reference for comparison. Table~\ref{tab:prediction_results} contains the performance metrics for the various models and proxies.

The baseline model achieved an \( R^2 \) of 0.54. While daytime population is a strong predictor of spectrum demand, it does not fully capture variations influenced by network deployment strategies and user mobility patterns, highlighting the need for additional features to better represent the spatial dynamics of spectrum demand.


  
The Deployed Bandwidth Proxy (\( P_{\text{BW}} \)) achieved an \( R^2 \) of 0.84, outperforming the baseline. This suggests that infrastructure deployment provides a meaningful approximation of spectrum demand. However, as observed in previous analyses, it may overestimate demand in areas with underutilized infrastructure or future-ready deployments.

The Active Users Proxy (\( P_{\text{Users}} \)) resulted in an \( R^2 \) of 0.68, indicating its ability to reflect end-user activity patterns. While this proxy incorporates user-driven network demand, it may underrepresent spectrum needs in sparsely populated regions where measurement data is limited.

The Combined Proxy (\( P_{\text{Combined}} \)) achieved the highest \( R^2 \) of 0.89, with the lowest Normalized RMSE and Normalized MAE. The results demonstrate that combining network deployment data with user-driven measurements enhances predictive accuracy by balancing infrastructure availability with actual user demand.



\begin{table}[ht]
\centering
\caption{Prediction Performance Metrics for Different Proxies}
\label{tab:prediction_results}
\resizebox{\linewidth}{!}{%
\begin{tabular}{l p{3cm} c c c}
\toprule
\textbf{Model} & \textbf{Proxy} & \textbf{\(R^2\)} & \textbf{Norm. RMSE} & \textbf{Norm. MAE} \\
\midrule
Baseline (Linear Reg.) &  (\(P_{\text{Combined}}\)) & 0.54 & 0.031 & 0.017 \\
\midrule
XGBoost &  (\(P_{\text{BW}}\)) & 0.84 & 0.026 & 0.016 \\
XGBoost &  (\(P_{\text{Users}}\)) & 0.68 & 0.027 & 0.018 \\
XGBoost &  (\(P_{\text{Combined}}\)) & 0.89 & 0.022 & 0.014 \\
\bottomrule
\end{tabular}%
}
\end{table}

\begin{figure}[ht]
\centering
\includegraphics[width=0.95\linewidth]{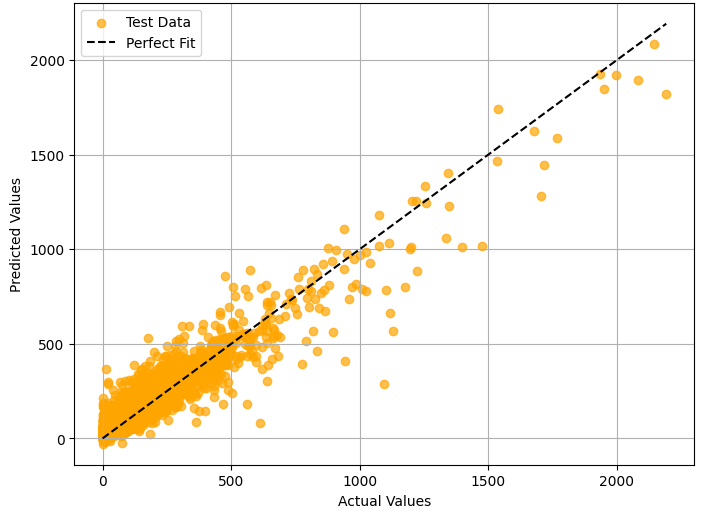}
\caption{Scatter plot of actual vs. predicted spectrum demand values for the Combined Proxy using the best-performing XGBoost model.}
\label{fig:scatter_pred_actual}
\end{figure}

To further illustrate the models effectiveness, Fig.~\ref{fig:scatter_pred_actual} shows a scatter plot of actual vs. predicted values for the Combined Proxy. The alignment between predicted and actual values demonstrates that this model provides a well-calibrated estimate of spectrum demand across different regions.

\begin{figure}[ht]
\centering
\includegraphics[width=0.98\linewidth]{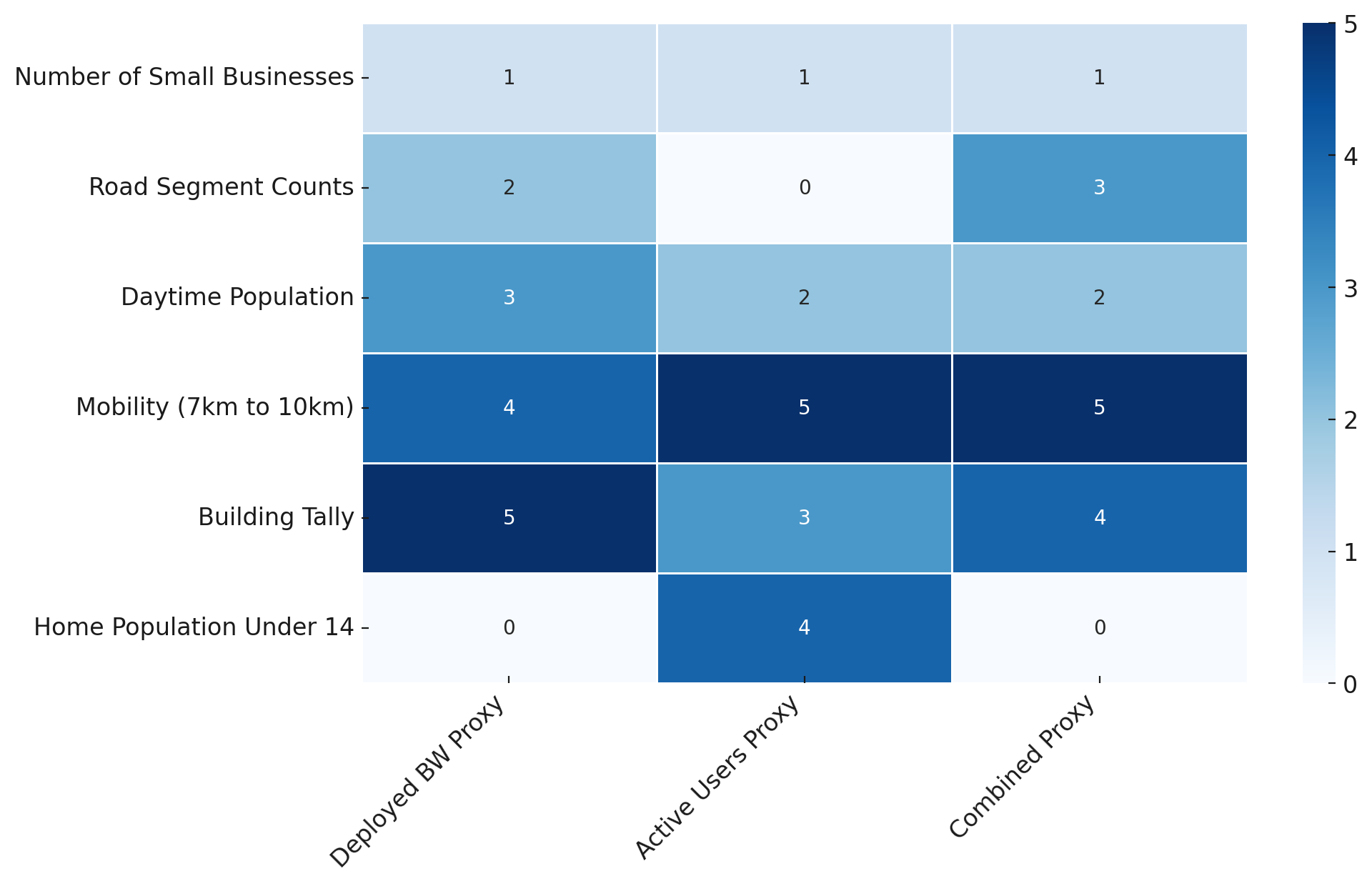}
\caption{Feature importance heatmap for XGBoost models trained with different proxies.}
\label{fig:feature_importance}
\end{figure}

Figure~\ref{fig:feature_importance} shows a heatmap of feature importance, focusing on the top five factors for each proxy model. Out of 21 features considered in the analysis, the number of small businesses is the most significant across all proxies. Road segment counts and daytime population also rank high, reflecting the link between network demand, commercial activity, and population. Mobility (trips between 7 km to 10 km) and building count further indicate the impact of movement and infrastructure density. Lastly, the home population under 14 years old is less influential, suggesting that areas with young families contribute less to network demand.

Overall, these results reinforce the validity of the Combined Proxy not only in direct validation against MNO traffic data but also in predictive modeling. 

\section{Conclusion}\label{sec:conclusion}
This study presented an approach to forecasting spectrum demand using AI techniques. The methodology integrated data from site licenses and crowdsourced sources, validated against real-world mobile network traffic data. An enhanced proxy was proposed that achieved an R$^2$ value of 0.89, indicating strong predictive accuracy. The ML models were tested and validated across five major Canadian cities, demonstrating their generalizability and robustness. As a result, this framework supports dynamic spectrum planning, enabling efficient and responsive resource allocation and policy adjustments to meet evolving network demands.


\bibliographystyle{ieeetr}
\bibliography{biblio}

\end{document}